\documentclass[journal,a4paper]{IEEEtran}
\usepackage{amsmath}
\usepackage{graphicx}
\usepackage{siunitx}
\usepackage{subcaption}
\usepackage{comment}
\usepackage{booktabs}
\usepackage{cite}
\usepackage{multirow}
\usepackage{graphicx}
\usepackage{placeins}
\usepackage{threeparttable}
\usepackage{tabularx} 
\usepackage{placeins}
\usepackage{amssymb}
\usepackage{url}
\usepackage[labelsep=period]{caption}

\title{Janus Metasurface Breaking Polarization Symmetry: Surface-Modulated Electromagnetic Wave Radiation with Coexistent Linear and Circular Polarization}

\author{
    Aparna Parameswaran,
    Hoyoung Kim,
    and Sangkil Kim,~\IEEEmembership{Senior Member, IEEE}%
    
    \thanks{A. Parameswaran, H. Kim and S. Kim are with Electrical and Electronics Engineering, Pusan National University, 46241, South Korea  (e-mail: Ksangkil3@pusan.ac.kr). This work was supported by the National Research Foundation of Korea (NRF) Grant funded by the Korea Government (MSIT) under Grant RS-2023-00237172.}%
   
}

\begin{document}
\maketitle

\begin{abstract}
In this work, a Janus metasurface based tensor impedance holographic antenna (JHA) is proposed that simultaneously radiates linearly polarized (LP) and circularly polarized (CP) beams from a single aperture excited by a single feed. The proposed design introduces modified tensor impedance equations to significantly reduce cross-polarization at higher radiation angles. It demonstrates broadband operation bandwidth of 0.5 GHz while maintaining high circular polarization purity. The design methodology is verified using aperture field integration theory, ensuring that the impedance distribution produces the desired far-field radiation patterns. Prototypes of three variations of the holographic antenna are fabricated, validating its performance. The radiation characteristics of the proposed antenna make it an attractive choice for advanced broadband communication applications.

\end{abstract}


\section{Introduction}
Higher gain, interference mitigation, spatial filtering, and compact, scalable structures make array antennas an attractive choice for modern communication applications such as radar, satellite links, 5G wireless systems, and Internet of Things (IoT) \cite{balanis2016antenna}. Various types of array antennas exist, classified based on their operating principles and radiation characteristics. Among them, holographic antennas have gained significant popularity due to their ability to control and manipulate electromagnetic waves using the principles of holography and impedance boundary conditions \cite{fong2010scalar,pandi2015design,minatti2016synthesis}. This approach enables the reconstruction of desired wavefronts from guided surface waves while maintaining a low-cost, low-profile, and lightweight design. Unlike conventional reflectarray antennas, which are often bulky because of their requirement for an offset feed, holographic antennas offer a low-profile and cost-effective alternative with relatively simpler feeding structures.

There are many studies on tensor impedance holographic antennas in the literature. These studies mainly focus on techniques to improve the performance characteristics of holographic antennas, along with methods for their efficient modeling and simulation\cite{ghosh2025simplified,zhu2024c,yang2023frequency,tong2023integrated,hu2022holographic,wen2023low,minatti2014modulated,pandi2015design,casaletti2016polarized,kwon2021modulated}. Most of these designs are either single-beam or multi-beam antennas operating at a single frequency. Additionally, they mainly emphasize aspects such as stable performance under conformal conditions, beam steering, miniaturization, and polarization control. More recently, a dual-band, dual-circularly polarized (CP) tensor impedance antenna has been proposed, which can operate at two distinct frequencies without compromising its radiation properties \cite{wang2025dual}.  Although \cite{fong2010scalar} reports CP radiation at $\SI{45}{\degree}$, it is achieved using an edge-fed holographic antenna. In \cite{hu2022holographic}, tensor impedance holographic surfaces for multi-beam and multi-polarization radiation have been explored. They propose a multi-beam tensor impedance holographic surface operating at 20 GHz that radiates multiple  linearly polarized (LP) beams and CP beams using four monopole feeds based on  sub-regional impedance superposition and selective switching between multiple feed points to achieve flexible beam steering and polarization control. In \cite{ghosh2025simplified}, axially quad-sectored impedance-modulated metasurface antennas are introduced, wherein orthogonal phase modulation produces a CP beam, and out-of-phase modulation produces an LP beam. However, in their approach, CP and LP radiation are realized using separate structures for each polarization. It can be concluded that the possibility of improved polarization purity at higher radiation angles and simultaneous LP-CP radiations from a single aperture with a single feed has not been explored in the literature. This is an important functionality as it is particularly useful in radar systems where target identification can be done effectively utilizing polarization diversity and to ensure signal integrity in wireless communication where polarization diversity can be exploited to reduce facing.

In the proposed work, a Janus metasurface based tensor impedance holographic antenna (JHA) that breaks polarization symmetry is introduced that allows excellent control over the polarization of radiated fields using simple design equations. The structure is excited by a single monopole antenna while being capable of simultaneously radiating LP and CP beams over a wide frequency band of 0.5 GHz without compromising the desired radiation characteristics. Additionally, modified tensor impedance equations are developed that significantly reduce cross-polarization at higher radiation angles, addressing one of the major limitations observed in conventional tensor impedance holographic antennas. The design methodology is initially validated using the well-known aperture field integration theory, ensuring that the impedance distribution results in the desired far-field radiation patterns. Furthermore, three design variations of the JHA are implemented and analyzed, demonstrating stable performance in terms of beam shape, polarization purity, and wide-band operation.

\section{Design Methodology}

\begin{figure}[t]
	\centering
	\includegraphics[width=0.8\columnwidth,keepaspectratio]{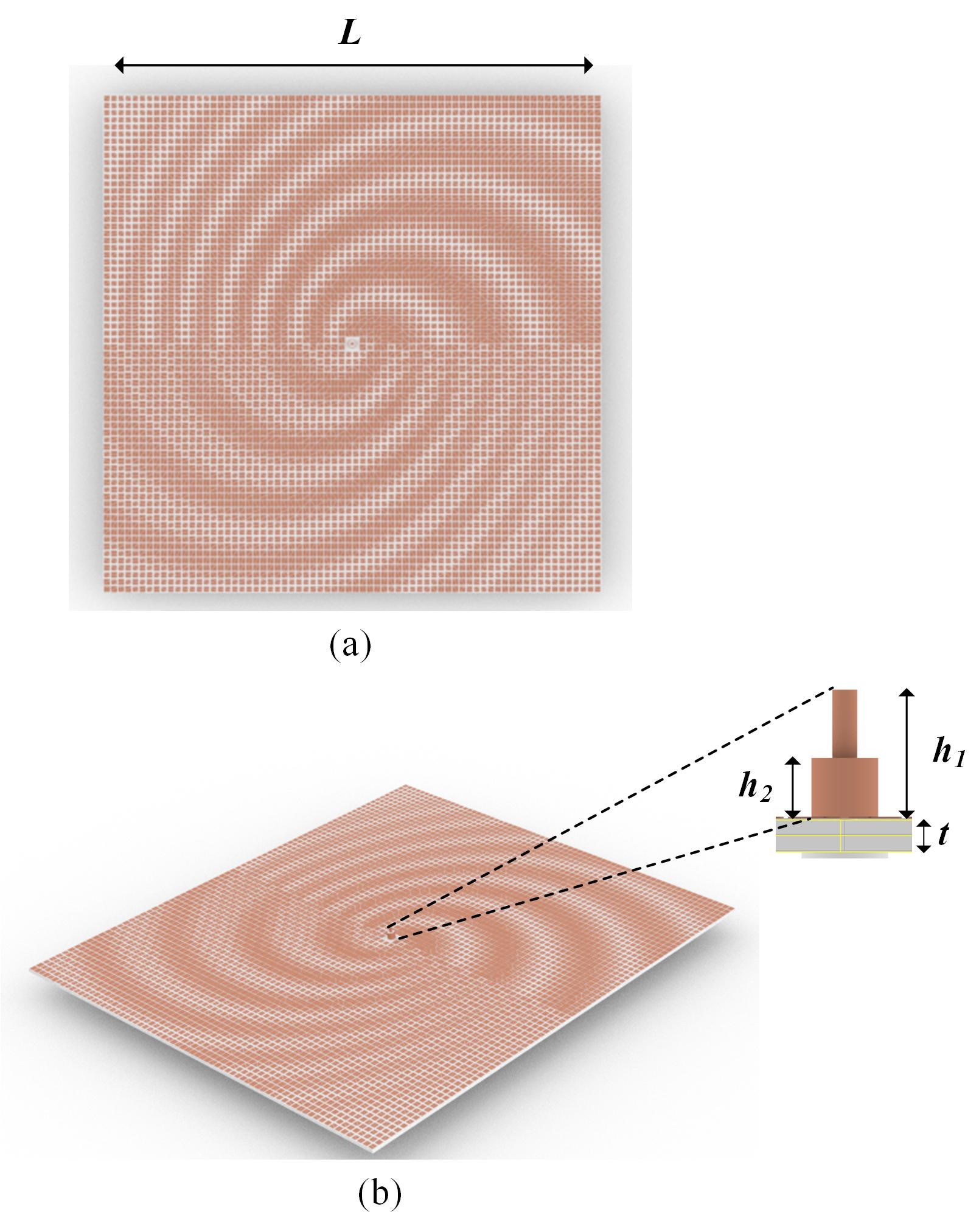}
	\caption{(a) Top view and (b) perspective view of the proposed dual-polarized JHA, with the inset in (b) showing the side view of the metallic-sleeved monopole antenna. The dimensions in mm are: $t$ = 1.57 mm, $h_{1}$ = 6.38 mm, $h_{2}$ = 3 mm, $L$ = 210 mm. }
     \label{fig1}
\end{figure}

Tensor impedance holographic antennas are widely studied due to their ability to provide enhanced gain and excellent control over the polarization of the radiated field. However, the possibility of simultaneously realizing LP - CP fields using a tensor impedance surface has not yet been explored in the literature. Therefore, the objective of this work is to address this gap by designing a tensor impedance holographic antenna capable of generating simultaneous LP and CP fields.

The geometry of the proposed JHA is shown in Fig. \ref{fig1}. It is implemented on a grounded 1.57 mm thick FR4 substrate ($\epsilon_{r}$ = 4.4 and tan $\delta$ = 0.02), patterned with sub-wavelength slotted metallic patches on the top surface. A monopole antenna placed at the center of the aperture, operating at 11.75 GHz and fed through a coaxial probe, excites the tensor surface to generate the desired radiation pattern. To achieve proper impedance matching between the probe and the antenna over a wide bandwidth, the monopole is enclosed partially in a metallic sleeve filled with a low-loss dielectric material (Teflon $\epsilon_{r}$ = 2.1 and tan $\delta$ = 0.001).

The hybrid electric and magnetic fields supported by the tensor impedance surface can be expressed as \cite{fong2010scalar}

\begin{equation}
E = E_{\mathrm{TM}} + \alpha E_{\mathrm{TE}}
\label{eq1}
\end{equation}

and, 

\begin{equation}
H = H_{\mathrm{TM}} + \alpha H_{\mathrm{TE}}
\label{eq2}
\end{equation}

where $\alpha$ denotes the degree of coupling between the TE and TM modes. The $E$ and $H$ fields represent the field components of a surface wave propagating in the direction of $k_t$.

These fields satisfy the impedance boundary condition
\begin{equation}
E={Z} J
\label{eq3}
\end{equation}
where $Z$ is the surface impedance tensor. To satisfy the principle of energy conservation, $Z$ is both reciprocal and anti-Hermitian. Substituting Eqs. \ref{eq1} and \ref{eq2} into Eq. \ref{eq3} yields the effective scalar impedance $Z_{eff}$ given by \cite{fong2010scalar,yao2019wide,meng2024anisotropic,yao2022comparisons}

\begin{equation}
\small
\begin{split}
\frac{k_z}{k} = & \frac{-j \left(Z_0^2 - Z_{xy}^2 + Z_{xx} Z_{yy}\right)}%
{2 Z_0 \left(Z_{yy} \cos^2\theta_k - Z_{xy} \sin 2\theta_k + Z_{xx} \sin^2\theta_k\right)} \\
& \pm \frac{1}{2 Z_0 \left(Z_{yy} \cos^2\theta_k - Z_{xy} \sin 2\theta_k + Z_{xx} \sin^2\theta_k\right)} \\
& \quad \times \Bigg[ 
\left(Z_0^2 - Z_{xy}^2 + Z_{xx} Z_{yy}\right)^2 \\
& \quad - 4 Z_0^2 \left(Z_{yy} \cos^2\theta_k - Z_{xy} \sin 2\theta_k + Z_{xx} \sin^2\theta_k\right) \\
& \quad \times \left(Z_{xx} \cos^2\theta_k + Z_{xy} \sin 2\theta_k + Z_{yy} \sin^2\theta_k\right)
\Bigg]^{1/2}
\end{split}
\label{eq4}
\end{equation}

\subsection{Extraction of the maximum effective scalar impedance $Z_{eff-max}$}
\begin{figure}[t]
	\centering
	\includegraphics[width=0.8\columnwidth,keepaspectratio]{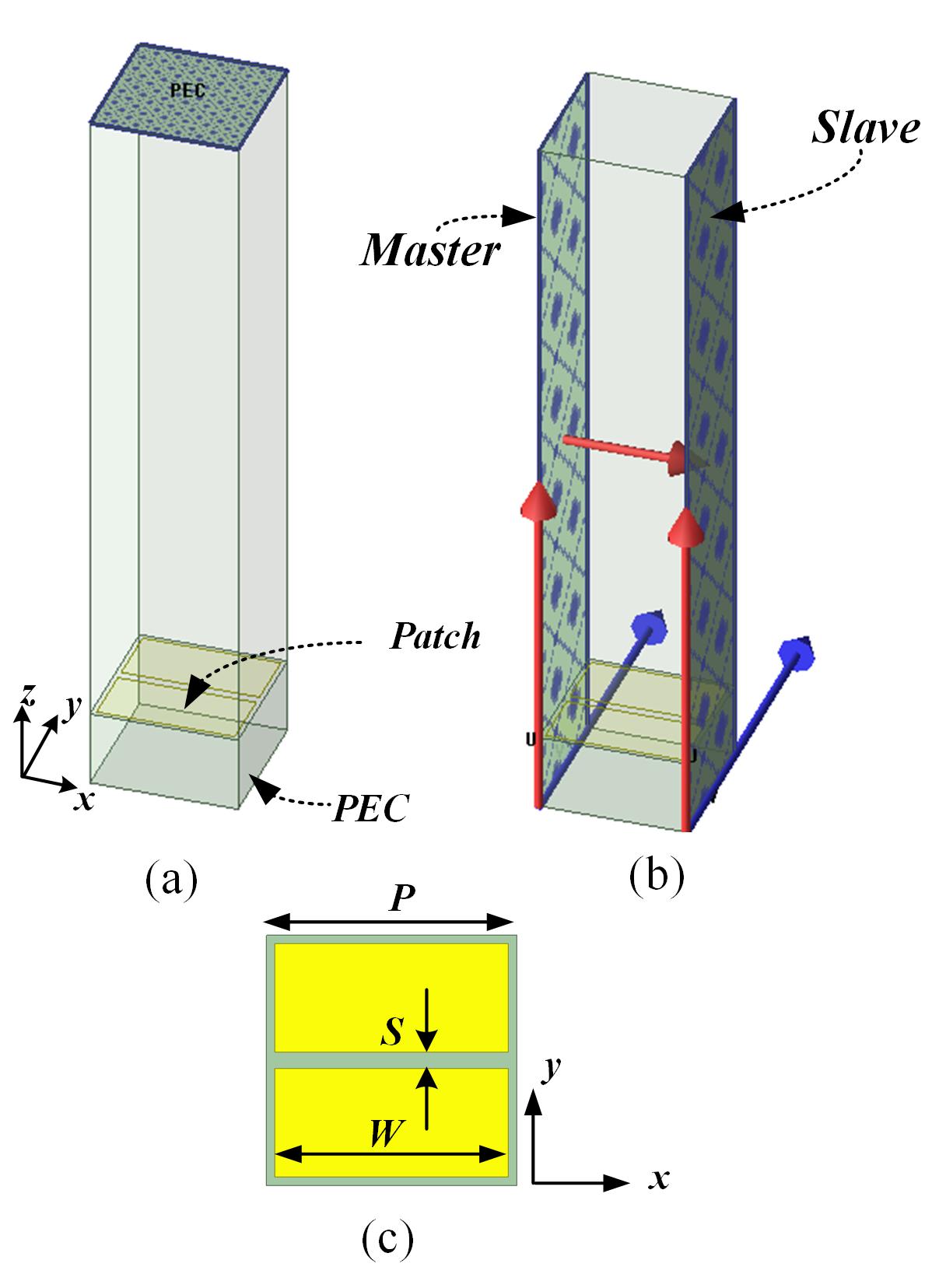}
	\caption{The proposed tensor unit cell with (a) PEC boundary condition, (b) periodic boundary condition and (c) top view.}
     \label{fig2}
\end{figure}

To simplify the design process, $Z_{eff}$ can be directly extracted as a function of the unit cell geometry, rather than computing the individual tensor impedance components from the geometry. The tensor unit cell employed in the design is shown in Fig. \ref{fig2}(c). It consists of a square metallic patch with a slot. The orientation of the slot determines the direction of maximum current flow and, consequently, the maximum effective scalar impedance, $Z_{eff-max}$. This orientation directly influences the values of the tensor impedance components $Z_{xx}$, $Z_{xy}$, and $Z_{yy}$. 

The unit cell is enclosed within a simulation volume, with periodic boundary conditions applied to the side walls (Fig. \ref{fig2}b) and PEC boundary conditions  (Fig. \ref{fig2}a)  assigned to the top and bottom surfaces. The height of the simulation volume is set to $\lambda/2$, corresponding to the center frequency of the antenna. The structure is then analyzed using the eigenmode solver in HFSS to extract the $Z_{eff-max}$ as a function of the gap length $g$ = $p - w$. For simplicity, the slot orientation is fixed along $\phi = \SI{0}{\degree}$ throughout the simulations. For each $g$, the phases $\phi_x$ and $\phi_y$ along both $x$ and $y$ directions can be computed for the desired operating frequency. From these values, the in-plane surface wave vectors can be found as 

\begin{equation}
\begin{aligned}
k_{x} = \frac{\phi_x}{p}\\
k_{y} = \frac{\phi_y}{p}
\end{aligned}
\end{equation}
where $p$ is the periodicity of the unit cell. The total surface wave vector is then computed as

\begin{equation}
k_t^2=k_x^2+k_y^2
\end{equation}

The normal wave vector $k_{z}$ is given by,
\begin{equation}
k_{z} = -j \sqrt{k_0^2 - k_t^2}
\end{equation}

Consequently, 

\begin{equation}
Z_{eff-max} = Z_{0}\frac{k_{z}}{k_{0}}
\end{equation}
where $k_{0}$ is the freespace wave number.

\begin{figure}[t]
	\centering
	\includegraphics[width=0.8\columnwidth,keepaspectratio]{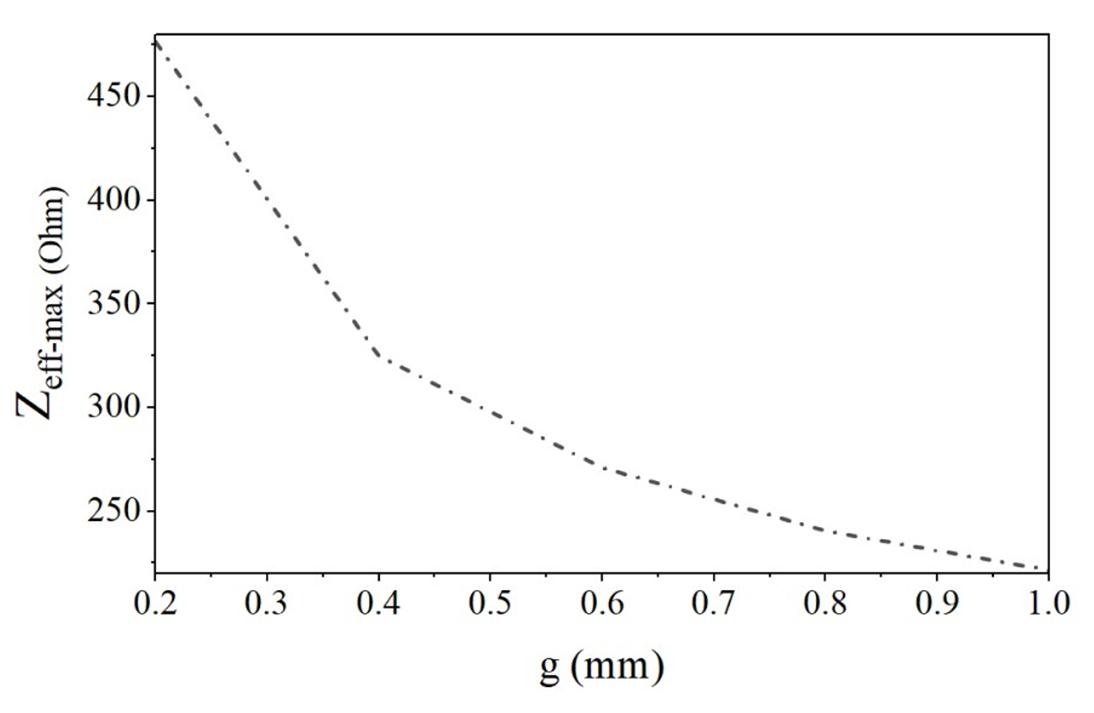}
	\caption{The variation of $Z_{eff-max}$ with respect to $g$ for the unit cell shown in Fig. \ref{fig2} with the slot along $\phi = \SI{0}{\degree}$.}
     \label{fig3}
\end{figure}

The relationship between $Z_{eff-max}$ and $g$, obtained from the full-wave simulation and the equations described above, is illustrated in Fig. \ref{fig3}. The results clearly show an inverse relationship, with $Z_{eff-max}$ decreasing with increasing $g$.

\subsection{Computing the tensor impedance components}
Using the boundary condition presented in Eq. \ref{eq3} and theory of optical holography \cite{dooley1965x,tricoles1977microwave}, the tensor impedance function can be derived as,

\begin{equation}
\small
Z(x) = j \begin{bmatrix} X & 0 \\ 0 & X \end{bmatrix} 
+ j \frac{M}{2} \, \mathrm{Im} \Big(
E_{\text{rad}}(x) \otimes J_{\text{surf}}^\textsuperscript{\dag}(x) 
- J_{\text{surf}}(x) \otimes E_{\text{rad}}^\textsuperscript{\dag}(x) 
\Big)
\label{eq8}
\end{equation}
where, ${E}_{r a d}$ is the desired radiated field and ${J}_{s u r f}({x})$ is the reference surface current excited by the source. This equation satisfies both the energy conservation principle and reciprocity requirements. Specifying ${E}_{r a d}$ and ${J}_{s u r f}({x})$ in Eq. \ref{eq8} would allow us to extract the in-plane tensor impedance components. 

The desired ${E}_{r a d}$ for LHCP radiation for the proposed design is

\begin{equation}
{E}_{\operatorname{rad}}(\mathbf{x})= (-j, 1,-j sin\theta_L) e^{-j {k}_0 x sin\theta_L}
\label{eq9}
\end{equation}
while for RHCP it is

\begin{equation}
{E}_{\operatorname{rad}}(\mathbf{x})= (-j, -1,-j sin\theta_L) e^{-j {k}_0 x sin\theta_L}
\label{eq10}
\end{equation}
where $\theta_L$ is the desired radiation angle.
The surface current generated by the source can be written as,

\begin{equation}
{J}_{s u r f}({x})=\frac{(x, y, 0)}{r} e^{-j k r}
\label{eq11}
\end{equation}

Substituting Eq.\ref{eq9}, Eq\ref{eq10} and Eq.\ref{eq11} into Eq. \ref{eq8} and extracting the impedance components gives us,

\begin{equation}
\begin{aligned}
Z_{x x} & =X-M \frac{x}{r} \cos \gamma \\
Z_{x y} & =Z_{y x}=\frac{M}{2}\left[-\frac{y}{r} \cos \gamma+\frac{x}{r} \sin \gamma\right] \\
Z_{y y} & =X - M \frac{y}{r} \sin \gamma
\label{eq12}
\end{aligned}
\end{equation}
for RHCP and ,
\begin{equation}
\begin{aligned}
Z_{x x} & =X-M \frac{x}{r} \cos \gamma \\
Z_{x y} & =Z_{y x}=\frac{-M}{2}\left[\frac{y}{r} \cos \gamma+\frac{x}{r} \sin \gamma\right] \\
Z_{y y} & =X - M \frac{y}{r} \sin \gamma
\label{eq13}
\end{aligned}
\end{equation}
for LHCP. Where X is the average surface reactance and M is the modulation depth. These values are chosen based on the maximum and minimum values of $Z_{eff-max}$ for the given range of $g$. The tensor impedance values for each unit cell can be determined based on the above equations to get the tensor impedance distribution on the aperture. Using the tensor impedance data, $Z_{eff-max}$ data obtained from the eigenmode solver and Eq. \ref{eq4}, the direction of $Z_{eff-max}$ for each unit cell can also be determined. The slot on the unit cell can then be cut along that direction in the actual design to maintain the desired chirality for LHCP or RHCP.

\subsection{Wide-band dual polarized JHA design}

The proposed JHA is capable of simultaneously radiating CP (LHCP) beam ($\phi = \SI{0}{\degree}$, +$\theta^o$) and LP beam ($\phi = \SI{180}{\degree}$, -$\theta^o$). Additionally, it preserves its radiation characteristics over a bandwidth of 0.5 GHz, extending from 11.5 GHz to 12 GHz.  

Holographic antennas are traditionally single-frequency antennas. However, wide-band operation can be achieved by taking the average of the tensor impedance distributions corresponding to the lower and upper cut-off frequencies of the desired band of operation \cite{wang2025dual}. 

\begin{equation}
Z(x,y) = \frac{1}{2} \sum_{i=1}^{2} Z_i(x,y)
\label{eq14}
\end{equation}
where $i = 1$ corresponds to $f_{l}$ = 11.5 GHz and $i = 2$ corresponds to $f_{u}$ = 12 GHz. Following the procedure outlined in detail in \cite{wang2025dual}, we can derive the final tensor impedance distribution on the aperture as,

\begin{equation}
\scriptsize
Z(x,y) =
\begin{cases}
\begin{bmatrix}
\dfrac{1}{2}\!\left[X_1+X_2-\dfrac{x D c}{r}\right] &
-\dfrac{1}{4}\!\left[\dfrac{y  D c+x D s}{r}\right] \\[8pt]
-\dfrac{1}{4}\!\left[\dfrac{y  D c+x D s}{r}\right] &
\dfrac{1}{2}\!\left[X_1+X_2-\dfrac{y D s}{r}\right]
\end{bmatrix} 
\end{cases}
\label{eq15}
\end{equation}
where $D_{c}$ = $M_1 \cos \gamma_1+M_2 \cos \gamma_2$ and $D_{s}$ is $M_1 \sin \gamma_1+M_2 \sin \gamma_2$. This impedance distribution radiates a single LHCP beam at the desired radiation angle and preserves the radiation characteristics for the bandwidth ranging from 11.5 GHz to 12 GHz. 

This basic wide-band LHCP tensor impedance holographic antenna can be modified into a dual CP-LP structure by controlling the impedance on the two halves ($y > 0$ and $y < 0$) of the aperture separately. To get the desired aforementioned radiation characteristics, the aperture impedance distribution is defined as follows.

{\scriptsize
\begin{equation}
Z(x,y) =
\begin{cases}
\begin{bmatrix}
\dfrac{1}{2}\!\left[X_1+X_2-\dfrac{x D c}{r}\right] &
-\dfrac{1}{4}\!\left[\dfrac{y  D c+x D s}{r}\right] \\[4pt]
-\dfrac{1}{4}\!\left[\dfrac{y  D c+x D s}{r}\right] &
\dfrac{1}{2}\!\left[X_1+X_2-\dfrac{y D s}{r}\right]
\end{bmatrix}, & y>0 \\[8pt]
\end{cases}
\label{eq16}
\end{equation}
}

This gives LHCP at $\phi = \SI{0}{\degree}$, +$\theta^o$. To get LP at $\phi = \SI{0}{\degree}$, -$\theta^o$, the condition is derived as follows
\begin{equation}
Z(x,y) =
\begin{cases}
Z_{1}(x,y), & x \equiv 0 \pmod{2}, \; y \equiv 1 \pmod{2} \\[6pt]
Z_{2}(x,y), & x \equiv 1 \pmod{2}, \; y \equiv 0 \pmod{2} \\[6pt]
Z_{3}(x,y), & x \equiv 1 \pmod{2}, \; y \equiv 1 \pmod{2} \\[6pt]
Z_{4}(x,y), & x \equiv 0 \pmod{2}, \; y \equiv 0 \pmod{2} \\[6pt]
\end{cases}
\label{eq17}
\end{equation}
where $Z_{1}(x,y)$ and $Z_{2}(x,y)$ can be given by,

{\scriptsize
\begin{equation}
\begin{aligned}
Z_{1}= Z_{2} =
\begin{cases}
\begin{bmatrix}
\dfrac{1}{2}\!\left[X_1+X_2+\dfrac{x D c}{r}\right] &
\dfrac{1}{4}\!\left[\dfrac{y D c - x D s}{r}\right] \\
\\[-8pt]
\dfrac{1}{4}\!\left[\dfrac{y D c - x D s}{r}\right] &
\dfrac{1}{2}\!\left[X_1+X_2-\dfrac{y D s}{r}\right]
\end{bmatrix}, & y<0
\end{cases}
\end{aligned}
\label{eq18}
\end{equation}
}

and $Z_{3}(x,y)$ and $Z_{4}(x,y)$ are given by Eq. \ref{eq16}. 

\begin{figure}[t]
	\centering
	\includegraphics[width=1\columnwidth,keepaspectratio]{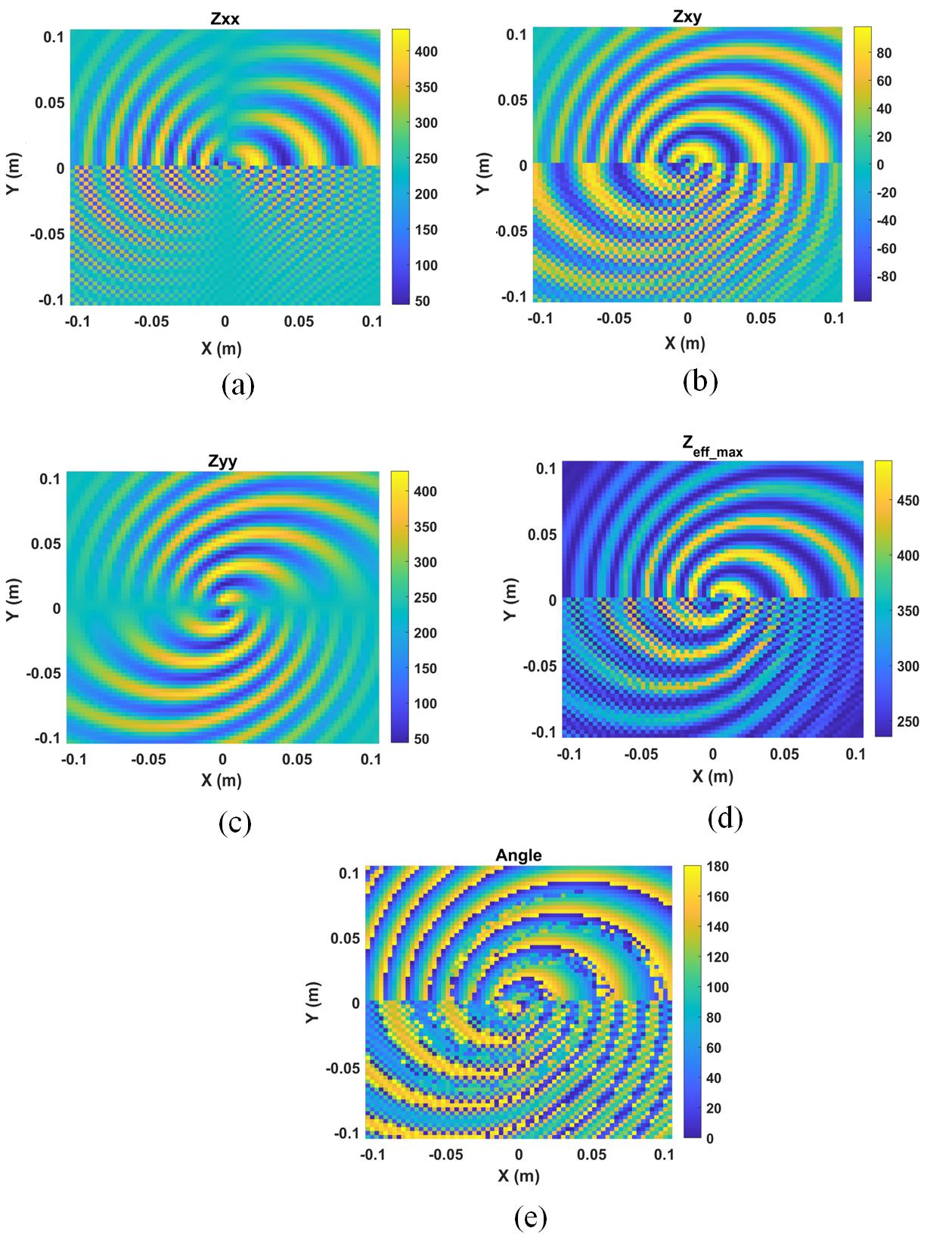}
	\caption{The tensor impedance distribution for the proposed dual-polarized wide-band holographic antenna (a) $Z_{xx}$, (b) $Z_{xy}$, (c) $Z_{yy}$, (d) $Z_{eff-max}$ and (e) angle distribution.}
     \label{fig4}
\end{figure}

The impedance distribution for the conditions outlined above at $f_{c}$ = 11.75 GHz for the radiation couplet $\phi = \SI{0}{\degree}$, $\theta$ = +$\SI{30}{\degree}$ : $\phi = \SI{180}{\degree}$, $\theta$ = -$\SI{30}{\degree}$ with $X_{1}$ = 0.64, $X_{2}$ = 0.56, $M_{1}$ = $M_{2}$ = 171, is shown in Fig. \ref{fig4}. 

\subsection{Analysis of the JHA using aperture field integration theory}
Aperture field integration (API) is an efficient technique that allows computation of far field radiation pattern analytically without having to solve rigorous and time-consuming Maxwell's equations \cite{balanis2016antenna,yaghjian2003equivalence,amini2020wide}.

 In this work, we use this technique to initially compute the aperture fields on the holographic antenna surface, followed by the prediction of the far field radiation pattern by using the 2D Fourier integrals of the aperture fields. This method is highly useful to gain physical insight into the  principle of achieving simultaneous LP and CP radiations from a single feed tensor impedance holographic antenna design and also to quickly predict its radiation characteristics.

 \begin{figure}[t]
	\centering
	\includegraphics[width=1\columnwidth,keepaspectratio]{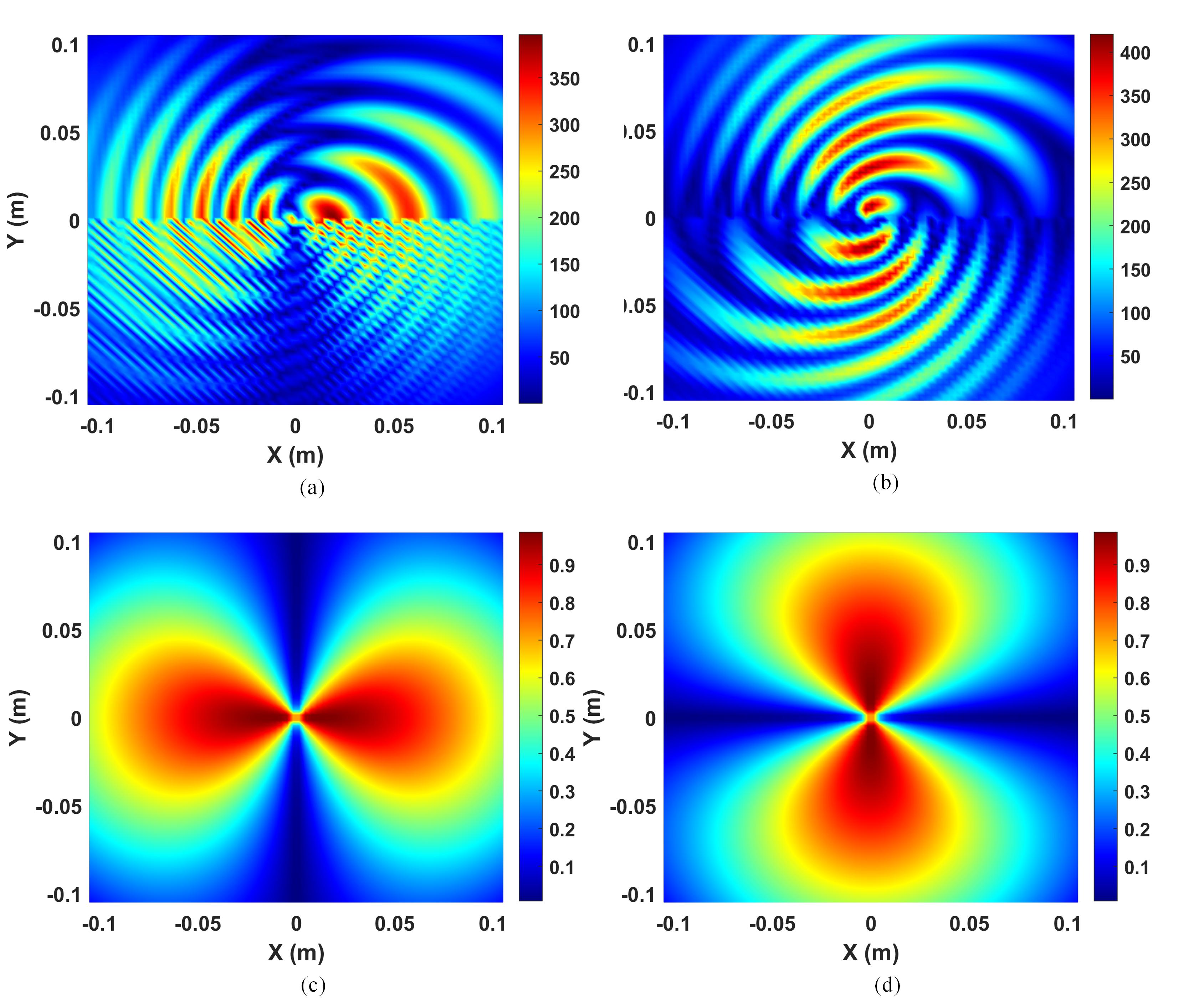}
	\caption{Calculated E-field  and current distribution using aperture integration theory: (a) Ex, (b) Ey, (c) Jx and (d) Jy.}
     \label{fig5}
\end{figure}

 Following the computation of the impedance distribution on the aperture as described in the previous section, the tangential fields on the aperture $E_x(x',y')$ and $E_y(x',y')$ can be obtained using the impedance boundary condition presented in Eq. \ref{eq3}. The computed electric field and current distributions are presented in Fig. \ref{fig5}.

 The spectral far fields $F_x(u, v)$ and $F_y(u, v)$ are computed next using 2D Fourier transform technique \cite{balanis2016antenna} as

\begin{equation}
F_x(u, v) = \iint\limits_{Aperture} E_x(x',y') \, e^{\,j k_0 \left( x' u + y' v \right)} \, dx' \, dy'
\label{eq19}
\end{equation}

\begin{equation}
F_y(u, v) = \iint\limits_{Aperture} E_y(x',y') \, e^{\,j k_0 \left( x' u + y' v \right)} \, dx' \, dy'
\label{eq20}
\end{equation}
where $u = sin \theta cos \phi$ and $v = sin \theta sin \phi$. While, $dx'$ and $dy'$ are the lattice periodicity of the antenna in the x and y directions.

From Eqs. \ref{eq19} and \ref{eq20}, the far field spatial components $E_\theta(\theta, \phi)$ and $E_\phi(\theta, \phi)$ can be computed as
\begin{equation}
\begin{aligned}
E_\theta(\theta, \phi) &= F_x(u,v) \, \cos\phi + F_y(u,v) \, \sin\phi \\
E_\phi(\theta, \phi)   &= \cos \theta (-F_x(u,v) \, \sin\phi + F_y(u,v) \, \cos\phi)
\label{eq21}
\end{aligned}
\end{equation}
Using $E_\theta$ and $E_\phi$ we can compute RHCP and LHCP patterns as a function of $\theta$ and $\phi$.

\begin{figure}[t]
	\centering
	\includegraphics[width=0.8\columnwidth,keepaspectratio]{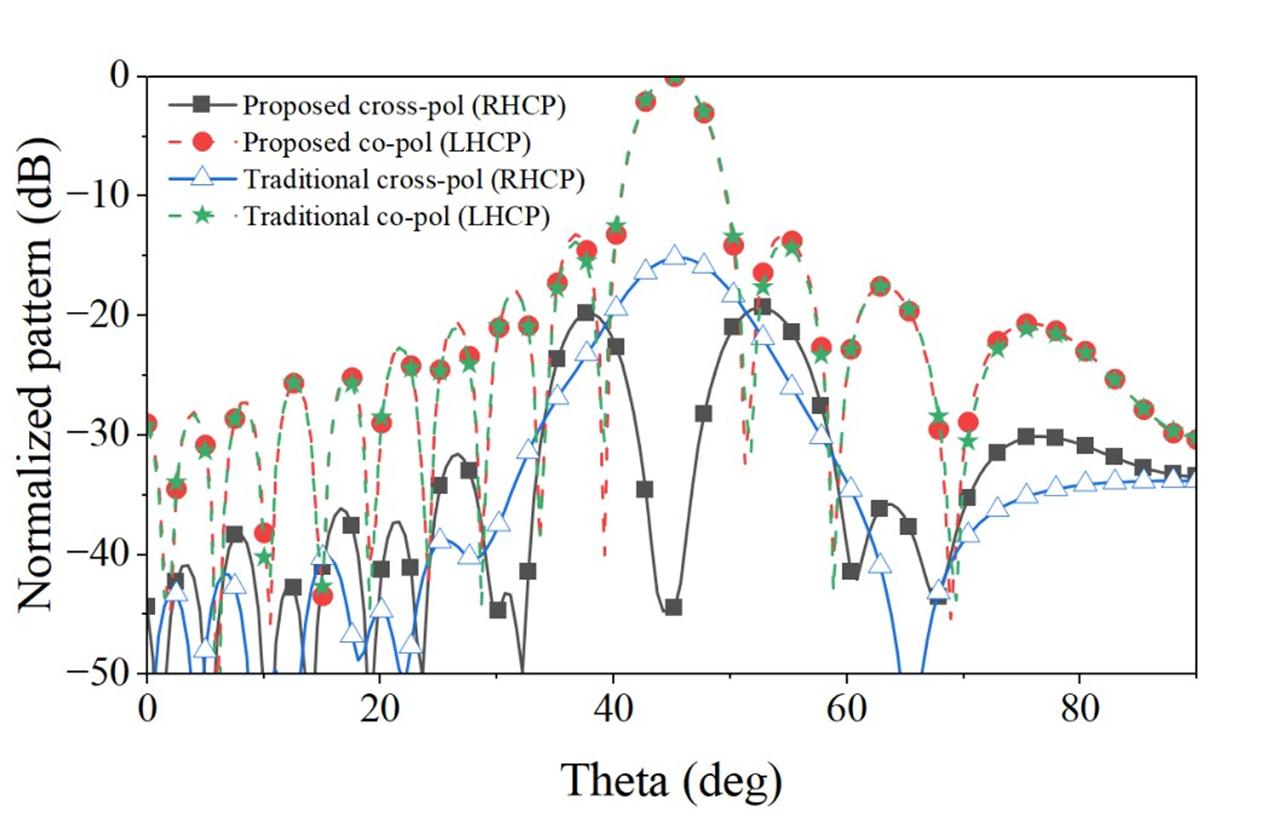}
	\caption{Comparison of the normalized co-pol and cross-pol radiation pattern of a single beam LHCP tensor impedance antenna obtained using traditional impedance equations \cite{fong2010scalar, wang2025dual} and the impedance equations used in this work at a radiation angle of $\SI{45}{\degree}$.}
     \label{fig6}
\end{figure}

To justify the choice of the impedance distribution used in this work and to demonstrate their cross-pol suppression capability, aperture field integration is used to compare  the co-pol and cross-pol radiation patterns of a single beam LHCP tensor impedance antenna obtained using traditional impedance equations and the impedance equations used in this work as presented in Fig. \ref{fig6}. It is well known that a tensor impedance holographic antenna is not suitable for higher radiation angles due to increased cross-pol degradation. This is due to the accumulation of phase errors on the surface of the aperture due to an increase in path length causing the projected tensor impedance to change. However, the choice of the desired $E_{rad}$ and consequently, the tensor impedance distribution can improve cross-pol suppression to an extent. It should be noted that in this work the $E_{rad}$ used (Eq. \ref{eq9} and \ref{eq10}) are different from the equations used in the literature \cite{fong2010scalar, wang2025dual}. The derived tensor impedance equations (Eq. \ref{eq12} and \ref{eq13}) are also different. These values were carefully chosen to ensure maximum cross-pol suppression for any given radiation angle. This can be clearly seen in Fig. \ref{fig6} for a radiation angle of $\SI{45}{\degree}$. The cross-pol predicted using the proposed equations is significantly lower than that predicted using traditional equations without degrading the co-pol pattern. Although a very effective analytical technique, aperture integration does not take into account near field interactions like edge diffraction or coupling between the elements. Therefore, the excellent cross-pol suppression we see in theory can be observed in full-wave simulation only up to a radiation angle of  $\SI{45}{\degree}$. Beyond this angle, we do observe higher cross-pol levels.

\begin{figure}[t]
	\centering
	\includegraphics[width=1\columnwidth,keepaspectratio]{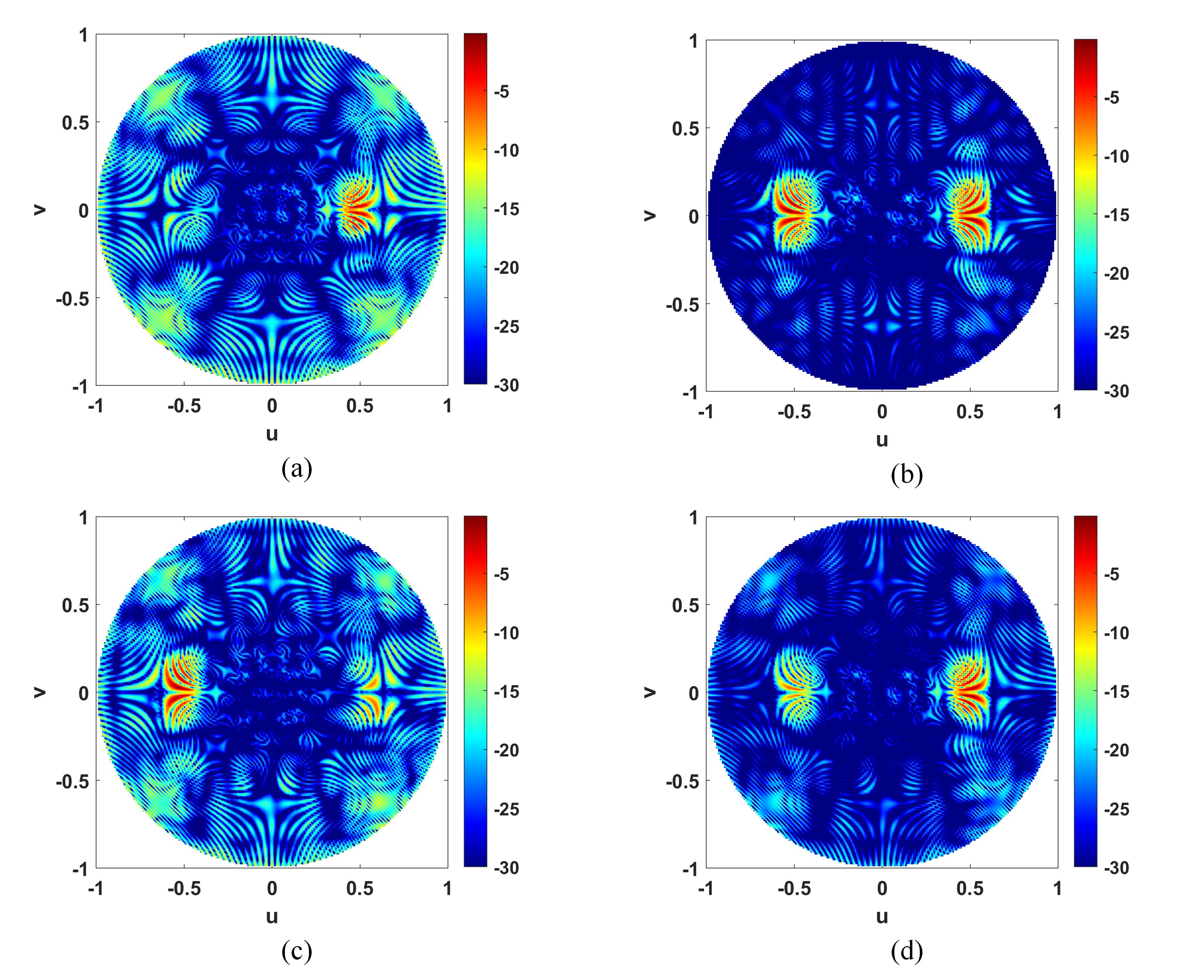}
	\caption{The spatial field magnitude of (a) $E_\theta$, (b) $E_\phi$, (c) $E_{RHCP}$ and (d) $E_{LHCP}$ projected onto the $u-v$ plane for the radiation couplet $\phi = \SI{0}{\degree}$, $\theta$ = +$\SI{30}{\degree}$ : $\phi = \SI{180}{\degree}$, $\theta$ = -$\SI{30}{\degree}$. }
     \label{fig7}
\end{figure}

\begin{figure}[t]
	\centering
	\includegraphics[width=0.78\columnwidth,keepaspectratio]{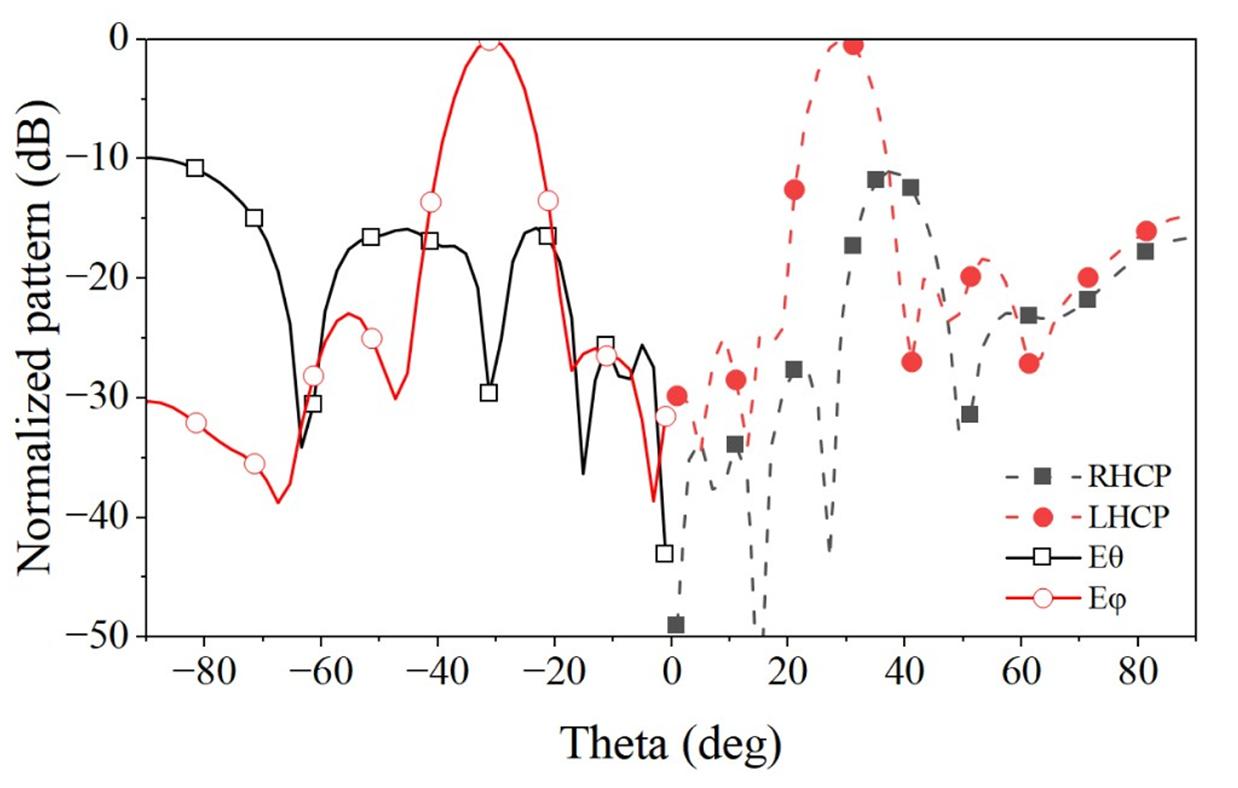}
	\caption{The computed circular and linear polarized radiation patterns using aperture integration technique for the radiation couplet $\phi = \SI{0}{\degree}$, $\theta$ = +$\SI{30}{\degree}$ : $\phi = \SI{180}{\degree}$, $\theta$ = -$\SI{30}{\degree}$. }
     \label{fig8}
\end{figure}

\begin{figure}[t]
	\centering
	\includegraphics[width=0.8\columnwidth,keepaspectratio]{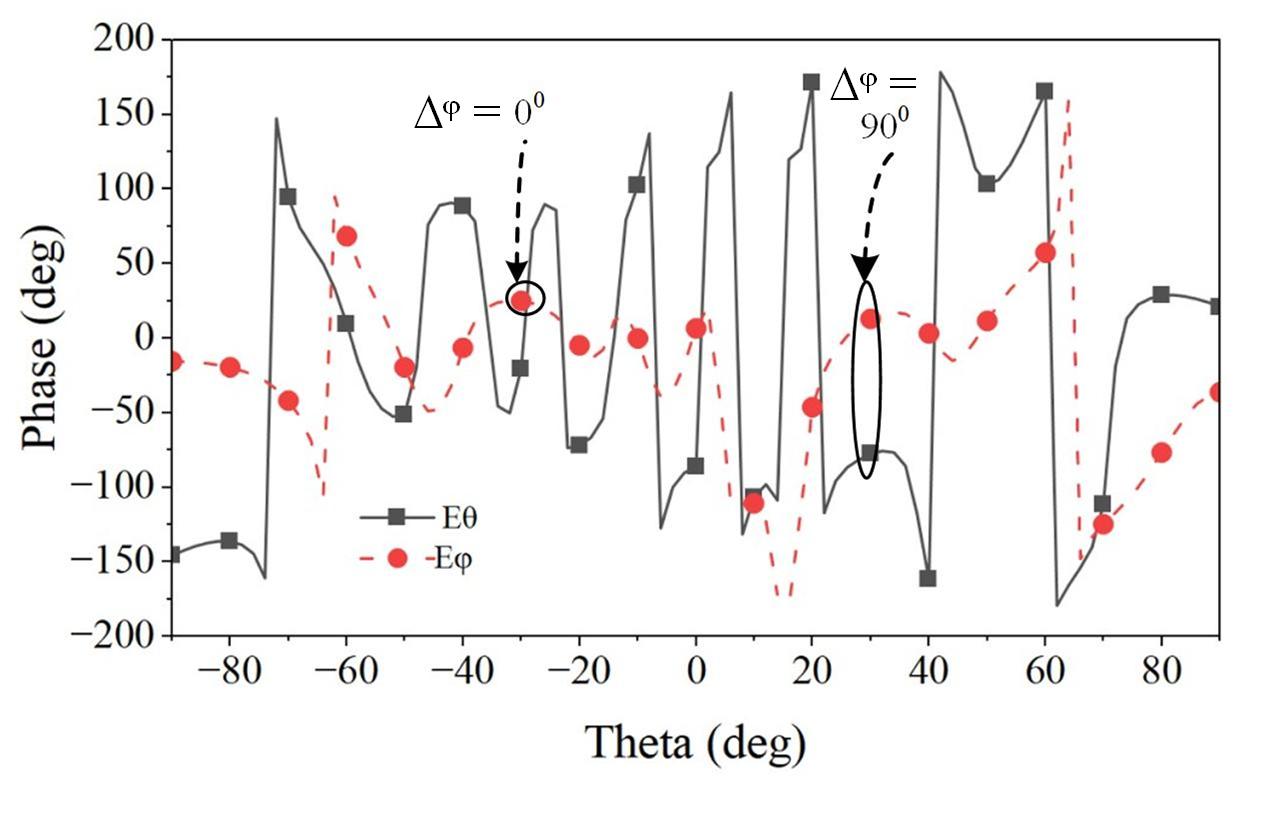}
	\caption{ The computed phase difference between the linear E field components of the radiation couplet $\phi = \SI{0}{\degree}$, $\theta$ = +$\SI{30}{\degree}$ : $\phi = \SI{180}{\degree}$, $\theta$ = -$\SI{30}{\degree}$. }
     \label{fig9}
\end{figure}

The magnitude of the spatial fields on $u-v$ plane for the proposed JHA computed using the aperture integration theory, corresponding to the impedance distribution presented in Fig. \ref{fig4}, is presented in Fig. \ref{fig7}. The 1D radiation pattern for $\phi = \SI{0}{\degree}$ cut for the proposed antenna is presented in Fig. \ref{fig8}. The phases of the linearly polarized E-field components are plotted in Fig. \ref{fig9}. Referring to Figs. \ref{fig7}, \ref{fig8} and \ref{fig9}, we can conclude that the tensor impedance structure efficiently ensures the following conditions:

\begin{itemize}
    \item For $y < 0$: $\angle E_\phi = \angle E_\theta $ : $|E_\phi| = |E_\theta|$
    \item For $y > 0$: $\angle E_\phi - \angle E_\theta = \SI{90}{\degree} $ : $|E_\phi| = |E_\theta|$
\end{itemize}

This means that for $y < 0$ $E_\theta$ and $E_\phi$ are directed along $\phi = \SI{180}{\degree}$, $\theta$ = -$\SI{30}{\degree}$. They interfere constructively in this direction while being in phase resulting in linear polarization. Conversely, for $y > 0$ $E_\theta$ and $E_\phi$ are directed along $\phi = \SI{0}{\degree}$, $\theta$ = $\SI{30}{\degree}$. They interfere constructively in this direction  while being out of phase by $\SI{90}{\degree}$ with $E_\phi$ leading $E_\theta$ resulting in LHCP polarization in this region. Because we have used the same periodicity for the sub-wavelength patches throughout the aperture, same amplitude for $E_\theta$ and $E_\phi$ is ensured. This can also be explained in terms of CP components. For $y > 0$, the LHCP and RHCP components are out of phase by $\SI{180}{\degree}$ resulting in CP (LHCP) radiation. For $y < 0$, LHCP and RHCP add up to produce LP ($E_\phi$) radiation.

Once the impedance distribution is finalized, using 2D interpolation technique, the inverse relationship between $Z_{eff-max}$ and $g$ is mapped to the 2D distribution presented in Fig. \ref{fig4} to realize the distribution of the slotted patch on the JHA geometry, where the direction of slots on each unit cell is determined using the data presented in Fig. \ref{fig4}(e). These distributions are then modeled in HFSS by linking MATLAB with HFSS using MATLAB-HFSS API.

\section{experimental verification}

\begin{figure}[t]
	\centering
	\includegraphics[width=1\columnwidth,keepaspectratio]{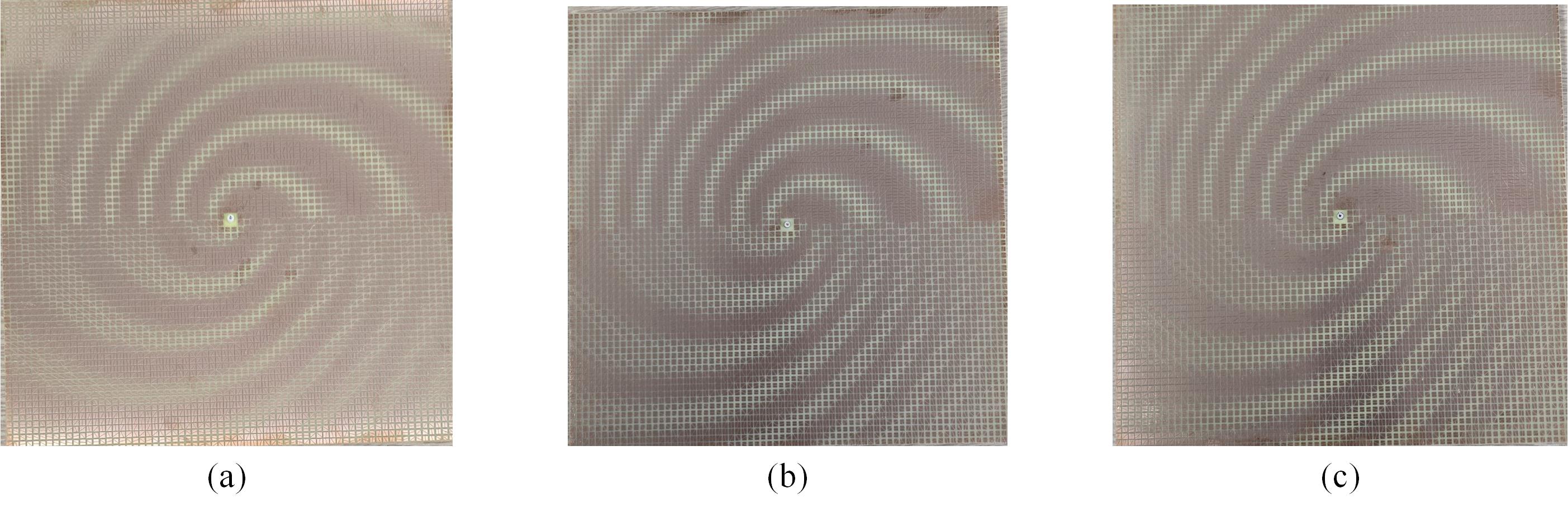}
	\caption{The top view of the fabricated prototypes for (a) Design I, (b) Design II and (c) Design III. }
     \label{fig10}
\end{figure}

\begin{figure}[t]
	\centering
	\includegraphics[width=1\columnwidth,keepaspectratio]{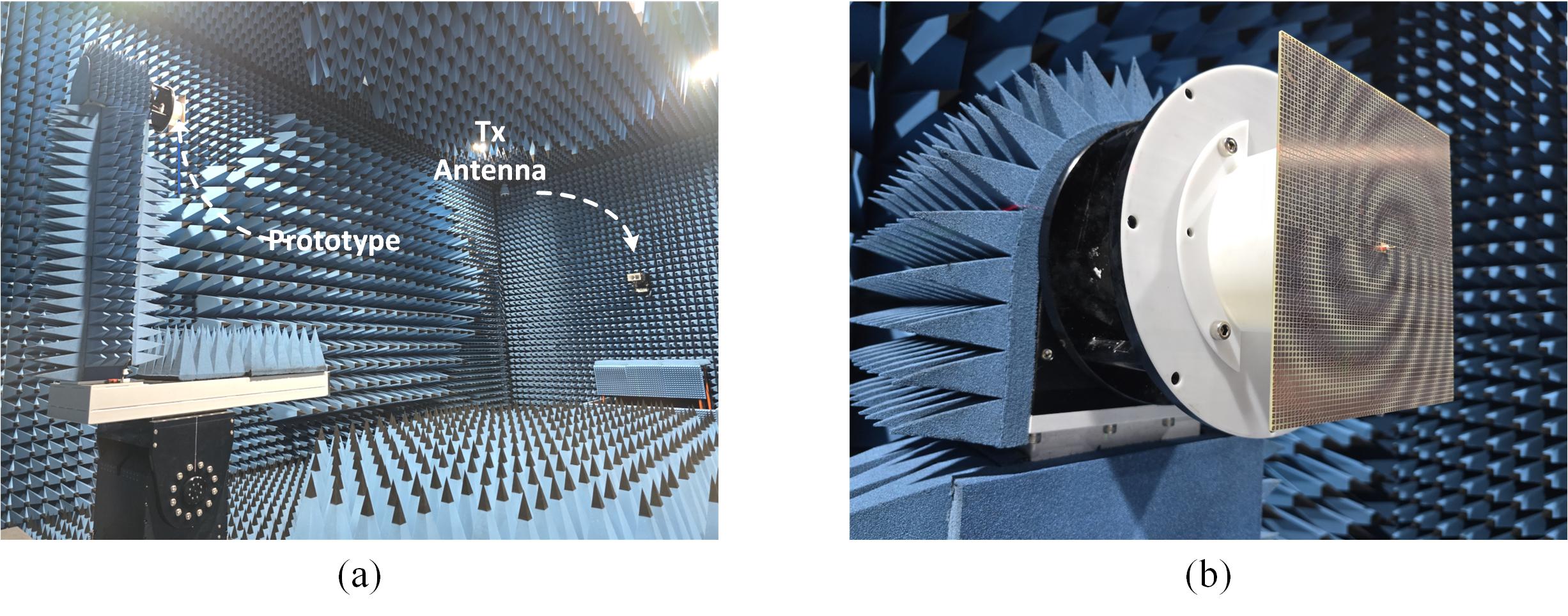}
	\caption{ (a) Far-field measurement setup and (b) prototype mounted on the rotating fixture.}
     \label{fig11}
\end{figure}

\begin{figure*}[htbp]
	\centering
	\includegraphics[width=1.7\columnwidth,keepaspectratio]{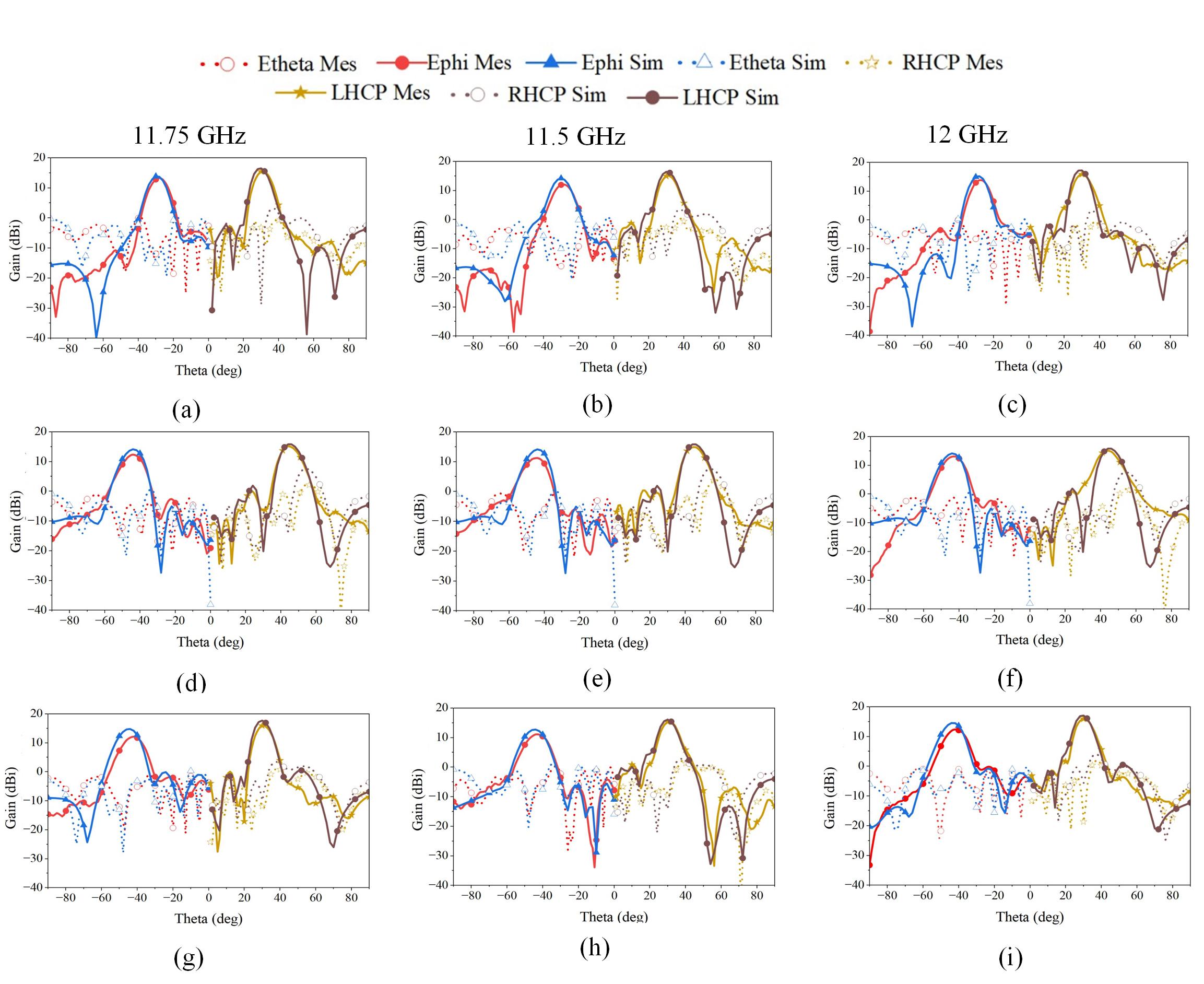}
	\caption{Comparison of simulated and measured gains obtained for Design I ((a),(b), and (c)), Design II ((d),(e), and (f))) and Design III ((g),(h), and (i)) at the $f_{l}$ = 11.5 GHz, $f_{c}$ = 11.75 GHz and $f_{u}$ = 12 GHz. }
     \label{fig12}
\end{figure*}

\begin{figure*}[htbp]
	\centering
	\includegraphics[width=1.7\columnwidth,keepaspectratio]{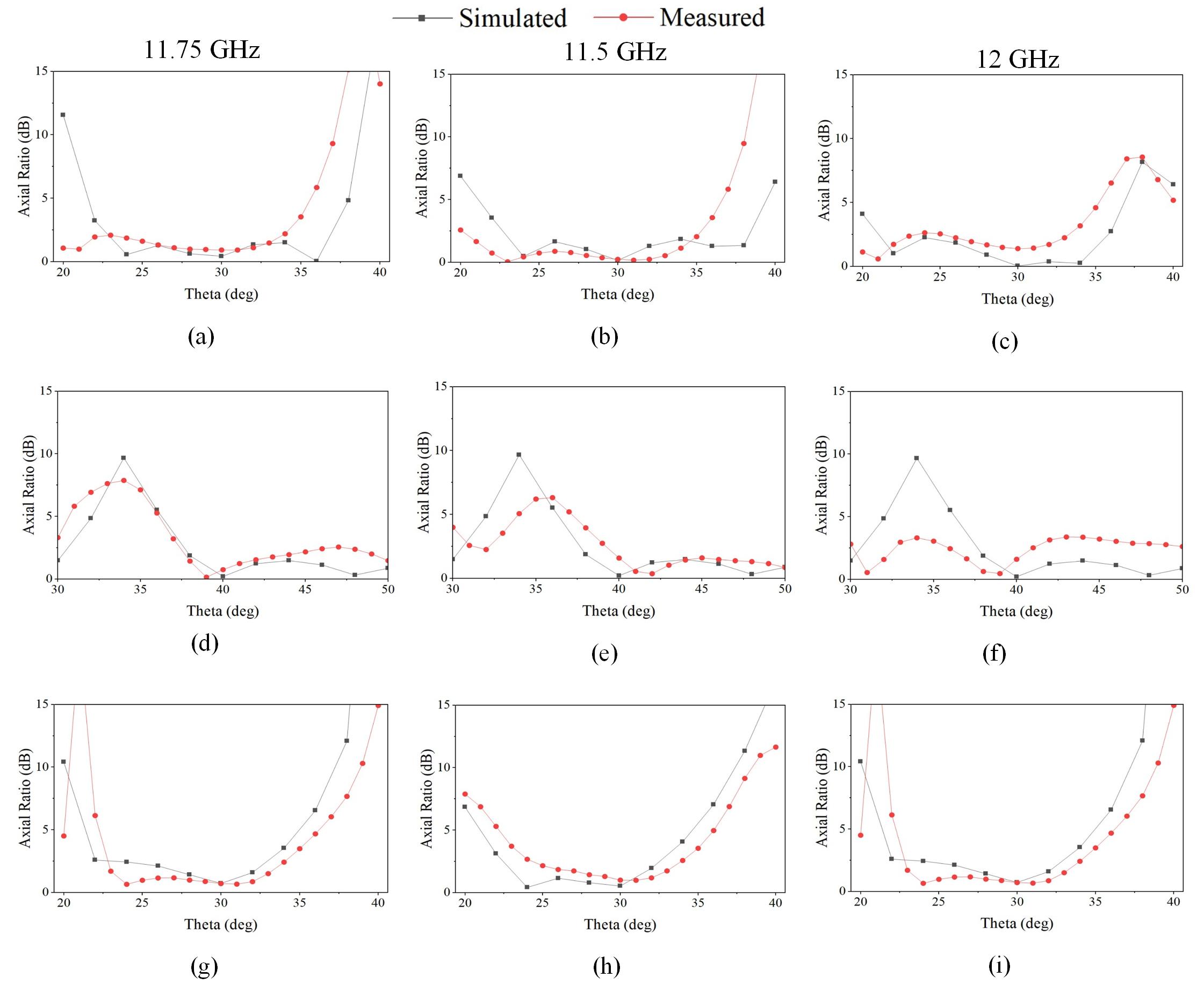}
	\caption{Comparison of simulated and measured axial ratios obtained for Design I ((a),(b), and (c)), Design II ((d),(e), and (f))) and Design III ((g),(h), and (i)) at the $f_{l}$ = 11.5 GHz, $f_{c}$ = 11.75 GHz and $f_{u}$ = 12 GHz. }
     \label{fig13}
\end{figure*}

\begin{figure}[t]
	\centering
	\includegraphics[width=1\columnwidth,keepaspectratio]{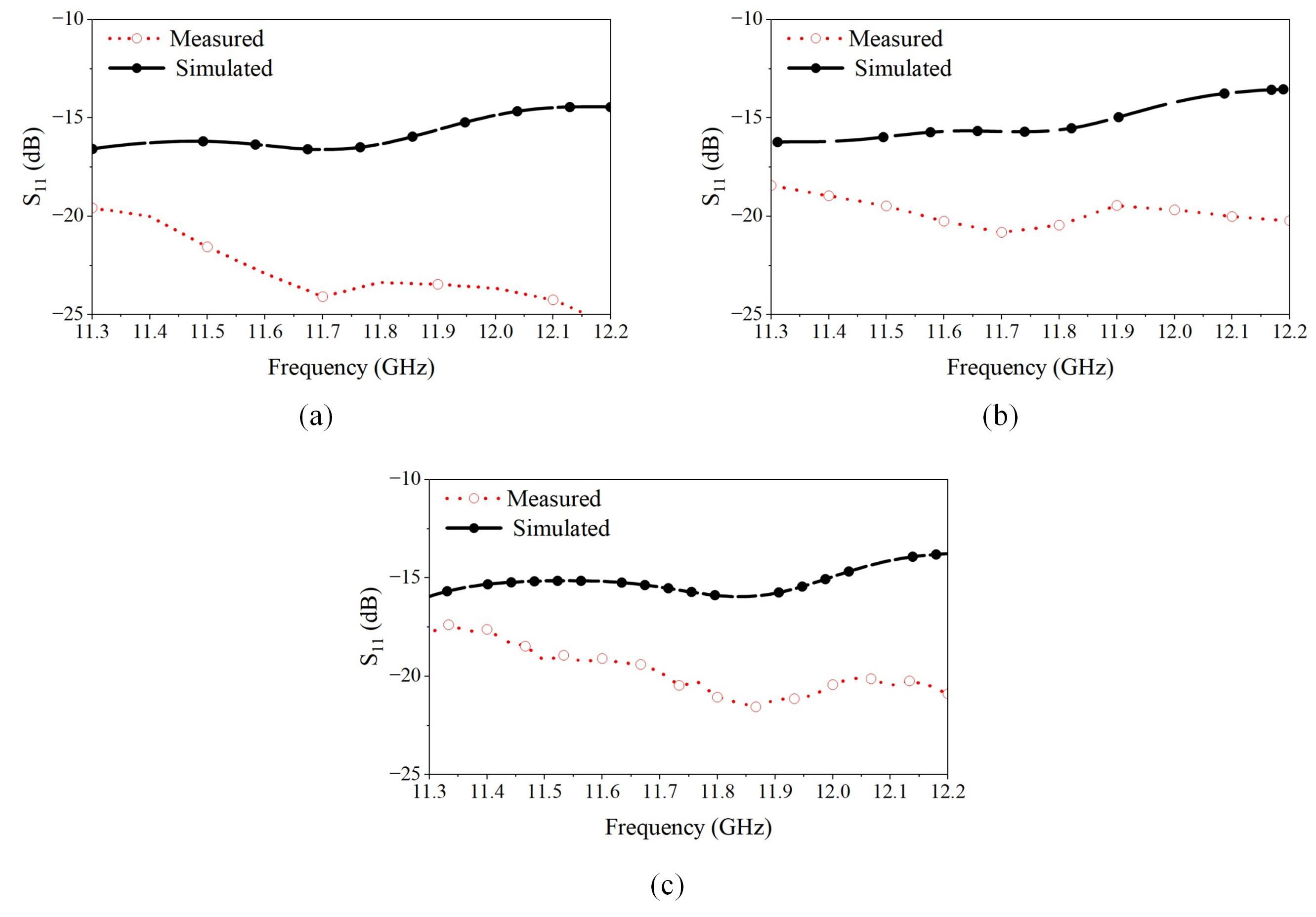}
	\caption{ Comparison of simulated and measured $|S_{11}|$ of (a) Design I, (b) Design II and (c) Design III.}
     \label{fig14}
\end{figure}

\begin{table*}[htbp]
  \centering
  \begin{threeparttable}
    \caption{Comparison of the proposed holographic antenna with the state-of-the-art presented in the literature.}
    \label{table1}
    \begin{tabularx}{\textwidth}{XXp{1.7cm}p{1.7cm}XXXXp{1.7cm}}
      \toprule
      Ref & Size (mm) & No. of ports & Bandwidth (GHz)& Axial Ratio (dB) & Maximum radiation angle & Simultaneous LP-CP capability & Dual-Beam & Gain \\
      \midrule
      \cite{fong2010scalar} & 406.40 $\times$ 254 & 1 (Edge-fed) & No & Not Reported & \SI{38}{\degree} & No & No & 21.80\textsuperscript{\dag} \\
      \\
      \cite{ghosh2025simplified} & 125 $\times$ 125 & 1 (Center-fed) & No & 1.25 & \SI{0}{\degree} & No & No & 19.25\textsuperscript{\dag} \\
      \\
      \cite{zhu2024c} & Not Reported & 1 (Edge-fed) & No & Not Reported & \SI{0}{\degree} & No & Yes & 11.80\textsuperscript{\dag}/31\textsuperscript{\dag} \\
      \\
      \cite{yang2023frequency} & 136 $\times$ 136 & 1 & No & Not Reported & \SI{0}{\degree} & No & Yes & 21.60\textsuperscript{\dag}/22.10\textsuperscript{\dag} \\
      \\
      \cite{tong2023integrated} & 191 $\times$ 191 & 1 (Edge-fed) & No & Not Reported & \SI{0}{\degree} & No & Yes & 26.80\textsuperscript{\dag}/28.20\textsuperscript{\dag} \\
      \\
      \cite{hu2022holographic} & Not Reported & 4 (edge-fed) & No & 0.90 & \SI{30}{\degree} & No & Yes & 17.90\textsuperscript{\dag}/22.5\textsuperscript{\dag} \\
      \\
      \cite{wen2023low} & 300 $\times$ 300 & 7 & No & Not Reported & \SI{36}{\degree} & No & Yes & 20\textsuperscript{\dag} \\
      \\
      \cite{wang2025dual} & 304 $\times$ 304 & 1 (Center-fed) & Yes (0.5 GHz) & 2.60 & \SI{30}{\degree} & No & Yes & 11.40\textsuperscript{\dag} /12.50\textsuperscript{\dag} \\
      \midrule
      {This work Design I} & 210 $\times$ 210 & 1 (Center-fed) & Yes (0.5 GHz) & 0.09 & \SI{30}{\degree} & Yes & Yes & 13.83\textsuperscript{\ddag} /16.48\textsuperscript{\dag} \\
      \\
      {This work Design II} & 210 $\times$ 210 & 1 (Center-fed) & Yes (0.5 GHz)& 0.59 & \SI{45}{\degree} & Yes & Yes & 14.11\textsuperscript{\ddag} /15.78\textsuperscript{\dag} \\
      \\
      {This work Design III} & 210 $\times$ 210 & 1 (Center-fed) & Yes (0.5 GHz) & 0.72 & \SI{30}{\degree} & {Yes} & Yes & 14.78\textsuperscript{\ddag} /17.71\textsuperscript{\dag} \\
      \bottomrule
    \end{tabularx}

    \begin{tablenotes}
      \footnotesize
      \item[\textsuperscript{\dag}] Gain in dBic \item[\ddag] Gain in dBi
      
    \end{tablenotes}

  \end{threeparttable}
\end{table*}

As a proof of concept, three variations of the JHA were fabricated and their performance was measured and compared to their simulated counterparts. Design I radiates LP along $\phi = \SI{180}{\degree}$, $\theta$ = -$\SI{30}{\degree}$ and LHCP along $\phi = \SI{0}{\degree}$, $\theta$ = $\SI{30}{\degree}$. Design II radiates LP along $\phi = \SI{180}{\degree}$, $\theta$ = -$\SI{45}{\degree}$ and LHCP along $\phi = \SI{0}{\degree}$, $\theta$ = $\SI{45}{\degree}$. Design III radiates asymmetrically with LP along $\phi = \SI{180}{\degree}$, $\theta$ = -$\SI{45}{\degree}$ and LHCP along $\phi = \SI{0}{\degree}$, $\theta$ = $\SI{30}{\degree}$. The design specifications like average surface reactance, modulation depth as well as the frequency of operation are the same for all the three samples. The fabricated prototype of each design is composed of 70 $\times$ 70 unit cells with a lattice constant of 3 mm. The photograph of all the three samples are presented in Fig. \ref{fig10}.

The setup used for the performance measurement of the prototypes is presented in Fig. \ref{fig11}. The setup is inside an anechoic chamber to avoid the undesirable influence of clutters from the environment. The transmitter antenna is a wide-band horn antenna with a maximum gain of 13 dBi. After calibrating the test setup, the prototypes were carefully mounted on a rotating fixture placed at the far-field distance from the horn antenna and their alignment was verified using the well known laser pointing technique. The radiation characteristics were measured for $\theta$ ranging from $\SI{-180}{\degree}$ to $\SI{180}{\degree}$ for $\phi = \SI{0}{\degree}$. A comparison of the radiation characteristics of the prototype of each design with its corresponding full-wave simulated results is presented in Fig. \ref{fig12}. The comparisons were done at lower cut-off frequency, upper cut-off frequency and the center frequency. It is observed that the measured and simulated results agree very well with each other at the cut-off and center frequencies thus validating the wide-band nature of the proposed design. It can also be seen that the results presented in Fig. \ref{fig12}(b) agree very well with the results predicted using the aperture integration theory presented in Fig. \ref{fig8}. A comparison of the axial ratios of the variants at the cut-off and center frequencies are also presented in Fig. \ref{fig13}. For Design I at $f_{c}$, $f_{l}$ and $f_{u}$, the axial ratio is consistently below 1 dB in full-wave simulations while the  measured values are below 2 dB. For Design II, the simulated axial ratios are below 1 dB while the measured values are below 2 dB. For Design III, both simulated and measured axial ratios are below 1 dB. These results validate not only the wide-band nature of the proposed holographic antenna but also its excellent polarization purity even for a radiation angle as high as  $\SI{45}{\degree}$. The comparison of the simulated and measured $S_{11}$ of the variants are presented in Fig. \ref{fig14} further ensures that good impedance matching is maintained within the bandwidth of interest.

The proposed JHA is compared to the state-of-the-art designs reported in the literature, as summarized in Table \ref{table1}. Compared to \cite{fong2010scalar,wang2025dual,wen2023low}, the proposed design is more compact while still achieving comparable gain, though being implemented on a highly lossy and dispersive substrate like FR4. Similar to \cite{ghosh2025simplified,wang2025dual,zhu2024c,yang2023frequency,tong2023integrated}, the antenna has a single center-fed port, however, among all the dual- and multi-beam structures, only the proposed design and \cite{wang2025dual} achieve this with a center-fed configuration. Furthermore, the proposed JHA exhibits the lowest simulated axial ratio among all designs considered for comparison, as well as the highest polarization purity at the highest radiation angle of $\SI{45}{\degree}$ where, a cross-polarization suppression as high as 23 dB is achieved. This is the first reported antenna capable of providing simultaneous LP–CP radiation along with wide-band operation at higher scan angles, while maintaining polarization purity ensured by the newly introduced modified tensor impedance equations. It can be concluded based on the comparison that the proposed JHA demonstrates strong potential for integration with next-generation wireless communication networks.

\section{conclusion}
In this paper, a wide-band, dual-polarized JHA is presented. The proposed design is capable of radiating both LP and CP waves simultaneously from a single aperture using only a single feed. Furthermore, each half of the aperture can radiate independently in asymmetric directions. Modified impedance distribution is proposed to ensure maximum cross-polarization suppression. The choice of tensor impedance design equations, as well as the underlying operating principle of the antenna, are validated using aperture field integration theory. To demonstrate proof of concept, three variations of the JHA are designed and fabricated, with good agreement observed between simulated and measured results. Owing to its ability to support simultaneous LP and CP radiation, it is a potential candidate for integration with radar systems, wireless communication networks and MIMO systems.

\bibliographystyle{ieeetran}
\bibliography{biblio1}

% Generated by IEEEtran.bst, version: 1.14 (2015/08/26)
\begin{thebibliography}{10}
\providecommand{\url}[1]{#1}
\csname url@samestyle\endcsname
\providecommand{\newblock}{\relax}
\providecommand{\bibinfo}[2]{#2}
\providecommand{\BIBentrySTDinterwordspacing}{\spaceskip=0pt\relax}
\providecommand{\BIBentryALTinterwordstretchfactor}{4}
\providecommand{\BIBentryALTinterwordspacing}{\spaceskip=\fontdimen2\font plus
\BIBentryALTinterwordstretchfactor\fontdimen3\font minus
  \fontdimen4\font\relax}
\providecommand{\BIBforeignlanguage}[2]{{%
\expandafter\ifx\csname l@#1\endcsname\relax
\typeout{** WARNING: IEEEtran.bst: No hyphenation pattern has been}%
\typeout{** loaded for the language `#1'. Using the pattern for}%
\typeout{** the default language instead.}%
\else
\language=\csname l@#1\endcsname
\fi
#2}}
\providecommand{\BIBdecl}{\relax}
\BIBdecl

\bibitem{balanis2016antenna}
C.~A. Balanis, \emph{Antenna theory: analysis and design}.\hskip 1em plus 0.5em
  minus 0.4em\relax John wiley \& sons, 2016.

\bibitem{fong2010scalar}
B.~H. Fong, J.~S. Colburn, J.~J. Ottusch, J.~L. Visher, and D.~F. Sievenpiper,
  ``Scalar and tensor holographic artificial impedance surfaces,'' \emph{IEEE
  {T}rans. {A}ntennas {P}ropag.}, vol.~58, no.~10, pp. 3212--3221, Oct. 2010.

\bibitem{pandi2015design}
S.~Pandi, C.~A. Balanis, and C.~R. Birtcher, ``Design of scalar impedance
  holographic metasurfaces for antenna beam formation with desired
  polarization,'' \emph{IEEE {T}rans. {A}ntennas {P}ropag.}, vol.~63, no.~7,
  pp. 3016--3024, Apr. 2015.

\bibitem{minatti2016synthesis}
G.~Minatti, F.~Caminita, E.~Martini, M.~Sabbadini, and S.~Maci, ``Synthesis of
  modulated-metasurface antennas with amplitude, phase, and polarization
  control,'' \emph{IEEE {T}rans. {A}ntennas {P}ropag.}, vol.~64, no.~9, pp.
  3907--3919, Jul. 2016.

\bibitem{ghosh2025simplified}
S.~Ghosh, P.~K. Mishra, and C.~Saha, ``Simplified theoretical characterization
  on polarized beam using axially quad-sectored impedance modulated metasurface
  antenna,'' \emph{IEEE {A}ntennas {W}irel. {P}ropag. {L}ett.}, vol.~24, no.~7,
  pp. 1704--1708, Feb. 2025.

\bibitem{zhu2024c}
J.~Zhu, S.~Liao, X.~Zhu, Y.~Yang, and Q.~Xue, ``C-/ka-band aperture-shared dual
  circularly polarized heterogeneous reflectarray for vehicular
  communications,'' \emph{IEEE Trans. Veh. Technol.}, vol.~73, no.~6, pp.
  8671--8680, Feb. 2024.

\bibitem{yang2023frequency}
W.~Yang, K.~Chen, J.~Zhao, T.~Jiang, and Y.~Feng, ``Frequency-multiplexed
  spin-decoupled metasurface for low-profile dual-band dual-circularly
  polarized transmitarray with independent beams,'' \emph{IEEE {T}rans.
  {A}ntennas {P}ropag.}, vol.~72, no.~1, pp. 642--652, Nov. 2023.

\bibitem{tong2023integrated}
X.~Tong, Z.~H. Jiang, Y.~Li, F.~Wu, J.~Wu, R.~Sauleau, and W.~Hong, ``An
  integrated dual-band dual-circularly polarized shared-aperture transmit-array
  antenna for k-/ka-band applications enabled by polarization twisting
  elements,'' \emph{IEEE {T}rans. {A}ntennas {P}ropag.}, vol.~71, no.~6, pp.
  4955--4966, Apr. 2023.

\bibitem{hu2022holographic}
L.-L. Hu, B.-Q. Li, X.-B. Liu, Y.-L. Tan, and C.~Zhu, ``Holographic impedance
  metasurface modulating multi-beam and polarization state,'' in \emph{2022
  International Conference on Microwave and Millimeter Wave Technology
  (ICMMT)}.\hskip 1em plus 0.5em minus 0.4em\relax IEEE, 2022, pp. 1--3.

\bibitem{wen2023low}
Y.~Wen, P.-Y. Qin, S.~Maci, and Y.~J. Guo, ``Low-profile multibeam antenna
  based on modulated metasurface,'' \emph{IEEE {T}rans. {A}ntennas {P}ropag.},
  vol.~71, no.~8, pp. 6568--6578, Aug. 2023.

\bibitem{minatti2014modulated}
G.~Minatti, M.~Faenzi, E.~Martini, F.~Caminita, P.~De~Vita,
  D.~Gonz{\'a}lez-Ovejero, M.~Sabbadini, and S.~Maci, ``Modulated metasurface
  antennas for space: Synthesis, analysis and realizations,'' \emph{IEEE
  {T}rans. {A}ntennas {P}ropag.}, vol.~63, no.~4, pp. 1288--1300, Dec. 2014.

\bibitem{casaletti2016polarized}
M.~Casaletti, M.~{\'S}mierzchalski, M.~Ettorre, R.~Sauleau, and N.~Capet,
  ``Polarized beams using scalar metasurfaces,'' \emph{IEEE {T}rans. {A}ntennas
  {P}ropag.}, vol.~64, no.~8, pp. 3391--3400, Aug. 2016.

\bibitem{kwon2021modulated}
D.-H. Kwon, ``Modulated scalar reactance surfaces for endfire radiation pattern
  synthesis,'' \emph{IEEE {T}rans. {A}ntennas {P}ropag.}, vol.~70, no.~1, pp.
  440--450, Jul. 2021.

\bibitem{wang2025dual}
P.~Wang, G.~Xu, B.~Yin, and W.~Wang, ``Dual-band dual-circularly polarized
  tensor holographic metasurface antenna for iov sensing and communications,''
  \emph{IEEE {I}nternet {T}hings {J}.}, vol.~12, no.~10, pp. 14\,447--14\,455,
  Jan. 2025.

\bibitem{yao2019wide}
M.~Yao, J.~L. Li, L.~Xia, and S.~S. Gao, ``Wide dual-band dual-circularly
  polarized holographic metasurface,'' \emph{J. Phys. D: Appl. Phys.}, vol.~52,
  no.~42, p. 425001, Aug. 2019.

\bibitem{meng2024anisotropic}
X.~Meng, H.~Zhang, T.~Wu, Y.~Li, A.~Zhang, L.~Ran, and X.~Chen, ``Anisotropic
  impedance holographic metasurface for near-field imaging,'' \emph{Photonics
  Res.}, vol.~12, no.~10, pp. 2226--2234, Sep. 2024.

\bibitem{yao2022comparisons}
M.~Yao, P.~Mei, G.~F. Pedersen, and S.~Zhang, ``Comparisons of scalar and
  tensor circularly-polarized holographic artificial impedance surfaces,'' in
  \emph{2022 16th European Conference on Antennas and Propagation
  (EuCAP)}.\hskip 1em plus 0.5em minus 0.4em\relax IEEE, 2022, pp. 1--5.

\bibitem{dooley1965x}
R.~Dooley, ``X-band holography,'' \emph{Proc. IEEE}, vol.~53, no.~11, pp.
  1733--1735, Nov. 1965.

\bibitem{tricoles1977microwave}
G.~Tricoles and N.~H. Farhat, ``Microwave holography-applications and
  techniques,'' in \emph{IEEE Proc.}, vol.~65, Jan. 1977, pp. 108--121.

\bibitem{yaghjian2003equivalence}
A.~Yaghjian, ``Equivalence of surface current and aperture field integrations
  for reflector antennas,'' \emph{IEEE {T}rans. {A}ntennas {P}ropag.}, vol.~32,
  no.~12, pp. 1355--1358, Dec. 1984.

\bibitem{amini2020wide}
A.~Amini, H.~Oraizi, M.~Hamedani, and A.~Keivaan, ``Wide-band polarization
  control of leaky waves on anisotropic holograms,'' \emph{Physical Review
  Applied}, vol.~13, no.~1, p. 014038, Jan. 2020.

\end{thebibliography}

\end{document}